\documentclass[12pt]{article}
\usepackage{fullpage}
\usepackage{caption}
\usepackage{graphicx} 
\usepackage{amsfonts}
\usepackage{amsmath,amsfonts,amsthm,bm}
\usepackage{amssymb}
\usepackage[colorinlistoftodos]{todonotes}
\usepackage{algorithm}
\usepackage{algpseudocode}
\usepackage{subfig}
\usepackage{appendix}
\newcommand{\probP}{\text{I\kern-0.15em P}}
\usepackage{color}
\usepackage{caption}
\usepackage{mathtools}
\usepackage{multirow}
\usepackage{float}
\usepackage{mathrsfs}
\usepackage{forloop}
\usepackage{url}
\usepackage{subfig}
\usepackage{authblk}
\usepackage{hyperref}

\author{Share\LaTeX}

\title{Clustering Heuristics for Robust Energy Capacitated Vehicle Routing Problem (ECVRP)}
\author{%
  Mark Pustilnik\footnote{Corresponding author: pkmark@berkeley.edu} , Francesco Borrelli \\ Department of Mechanical Engineering, University of California at
Berkeley, Berkeley, CA 94701 USA }
\date{February 2024}

\begin{document}

\maketitle

\begin{center} keywords: Robust ECVRP, Mixed Integer Programming, Clustering, Stochastic VRP, Vehicle Routing Problem, Electric Vehicles \end{center}

\section{ABSTRACT}

The paper presents an approach to solving the Robust Energy Capacitated Vehicle Routing Problem (RECVRP), focusing on electric vehicles and their limited battery capacity. 
A finite number of customers, each with their own demand, have to be serviced by an electric vehicle fleet while ensuring that none of the vehicles run out of energy. The time and energy it takes to travel between any two points is modeled as a random variable with known distribution.
We propose a Mixed Integer Program (MIP) for computing an exact solution and introduce clustering heuristics to enhance the solution speed. This enables efficient re-planning of routes in dynamic scenarios. The methodology transforms the RECVRP into smaller problems, yielding good quality solutions quickly compared to existing methods. We demonstrate the effectiveness of this approach using a well-known benchmark problem set as well as a set of randomly generated  problems.

\section{Introduction}
The Vehicle Routing problem (VRP) is a well-known combinatorial optimization challenge with significant practical implications in transportation and logistics management - Visiting $N$ nodes by $M$ cars such that every car starts and finishes at $K$ central nodes (Depots) and each node is visited a single time while minimizing a cost function. The cost function can be a simple total tour time, distance travelled by the vehicles or can be something more complex like the total operation cost (function of time and energy consumption). The ECVRP is an extension of the VRP - in addition to the constraints of the VRP problem, energy constraints are added. Every drive between customers takes some amount of energy that is subtracted from the vehicle's state-of-charge (SOC). The vehicle SOC must always be positive and, at the end of the tour, each truck has to return with enough energy so that it can be charged fully overnight for the next day tour. There are Charging Stations (CS) that can be visited during the tour to charge the vehicle battery with a known, possibly nonlinear, charging profile. The cost function to minimize is a function of the total time travelled by all vehicles and total charging time and can be other operational costs. In addition, each customer has some demand to be fulfilled by the vehicle and the sum of demands of all customers in a single tour can't exceed the capacity of the vehicle. Additional constraint may be added such as a limit on the maximal number of nodes visited by a single vehicle. The ECVRP, similarly to the classic VRP problem, is NP-hard and is impractical to solve to optimally for large scale problems \cite{article2}. There are many local methods for solving this problem and a literature review goes beyond the scope of this paper (\cite{article1,mavrovouniotis2020benchmark}). The solution to these problems is a tour - a set of nodes for each vehicle to visit and their order. 
This paper focuses on an important extension of the ECVRP which considers the stochastic nature of the two key parameters defining the problem: the time and the energy it takes to travel between nodes. The solution of the routing problem should take into account the probability associated with each tour both in the cost function and in the constraints. We will assume that the time and the energy associated to an edge is normally distributed. We first formulate a problem with a stochastic cost and and  chance-constrains, and then transform the problem into a deterministic MIP.

The main contribution of this paper is as follows: 1) formulate the  RECVRP problem as
a Mixed Integer Program (MIP), 2) develop a simple heuristic method to solve the RECVRP problem, by first clustering the data into $M$ groups, and then solving a smaller robust energy travelling salesman (RECTSP) problem. RECTSP is the problem of assigning the order of visit of customers while meeting the energy and load and minimizing the probabilistic tour time. 
We emphasize that the primary contribution of this paper is an extremely simple algorithm for solving the RECVRP Problem. According to tests on benchmark examples, this algorithm can generate high-quality tours in a short amount of time, comparable to those produced by state-of-the-art solvers.

\begin{figure}
    \centering
    \subfloat[\centering Tour Map and SoC Prediction] {{\label{fig:tour_example:a}\includegraphics[width=7cm]{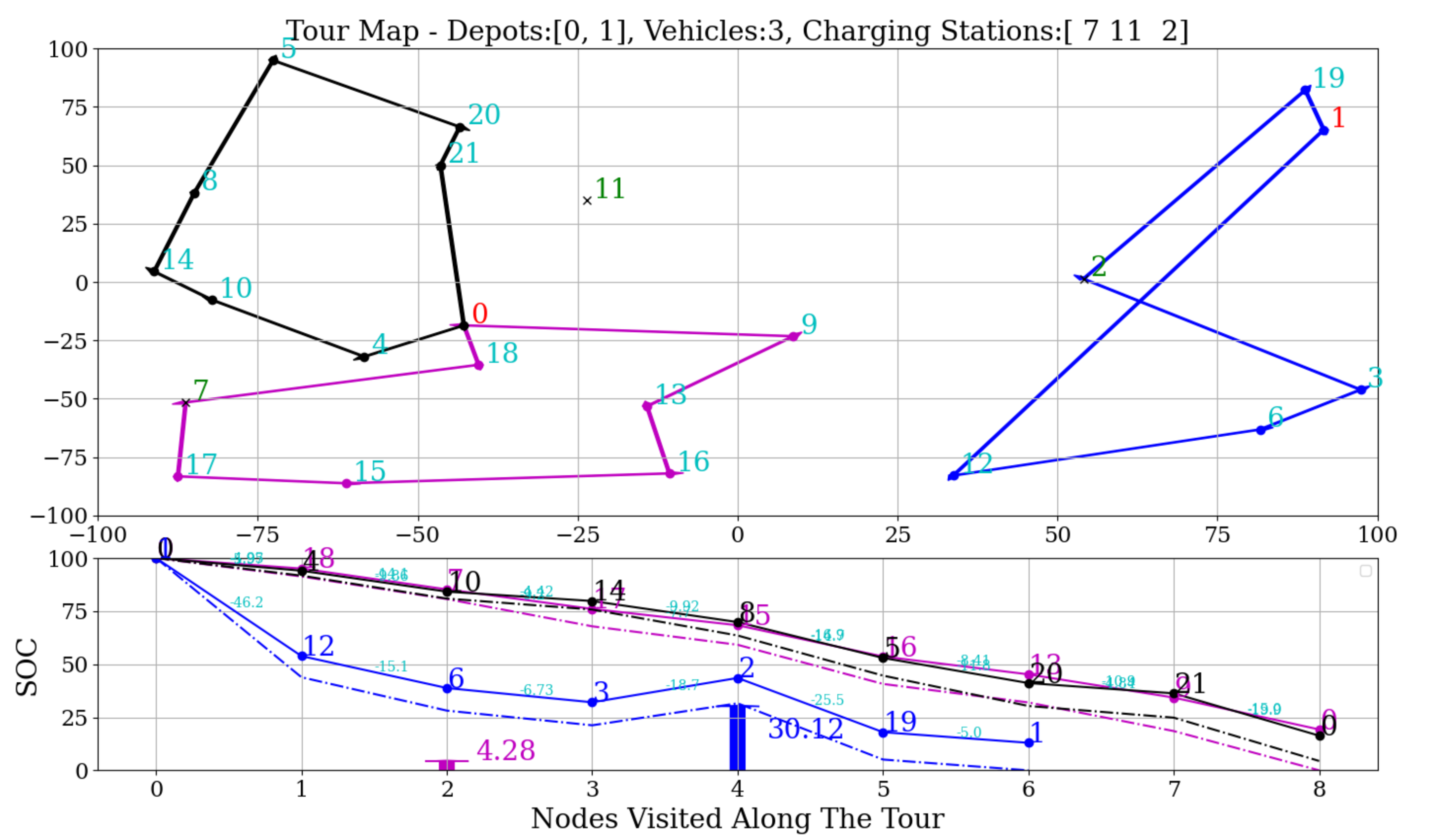} }}
    \qquad
    \subfloat[\centering MC of Time and SoC Statistics]{{\label{fig:tour_example:b}\includegraphics[width=7cm]{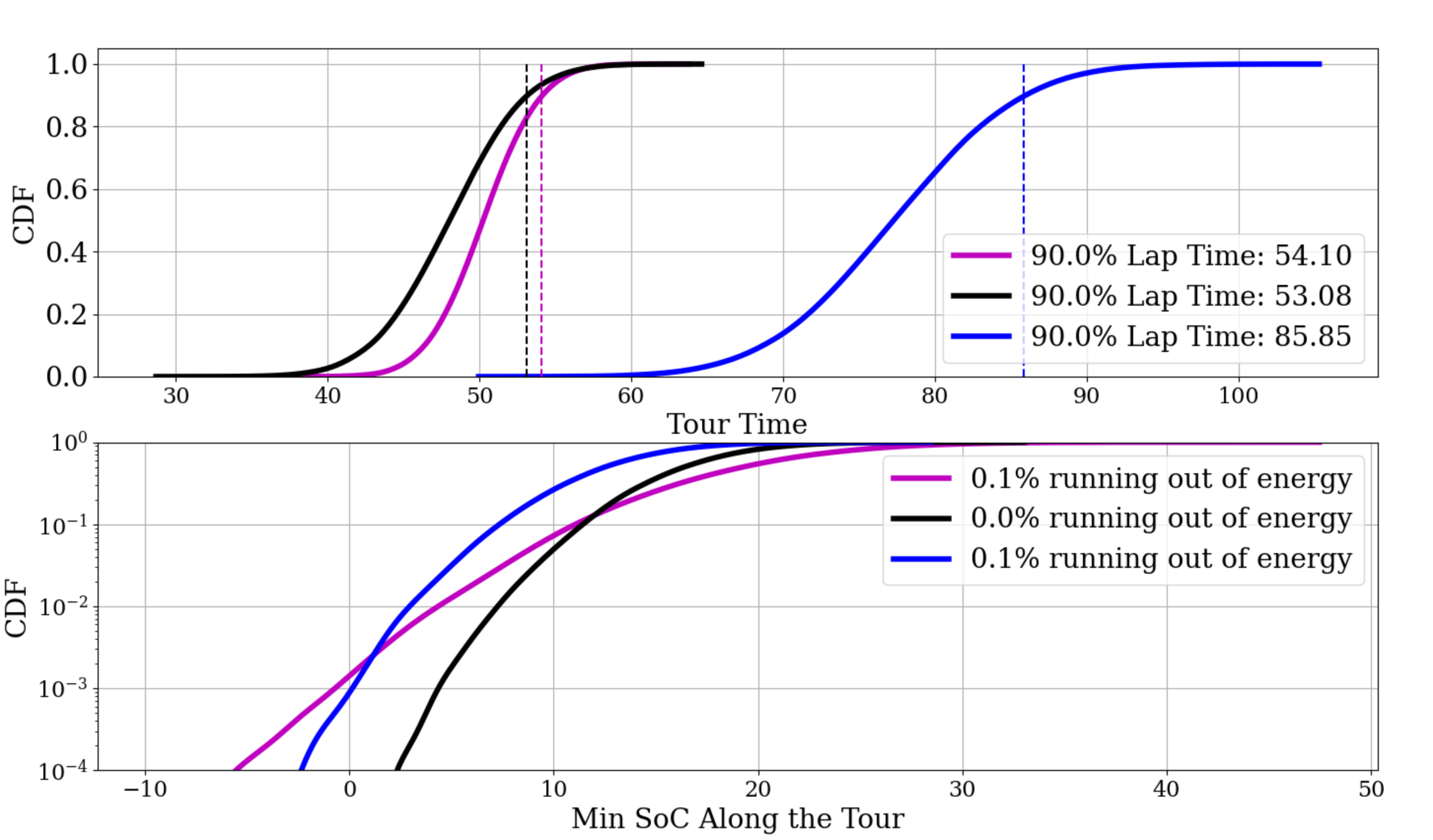} }}
    \caption{Tour Example}
    \label{fig:tour_example}
\end{figure}

Figure \ref{fig:tour_example} shows an example of a RECVRP tour solved using the MIP formulation with Gurobi solver \cite{gurobi} presented in section \ref{MIP}. It shows the nodes on a graph and the tour of 3 vehicles leaving 2 depots. Figure \ref{fig:tour_example:a} also shows the predicted average State of Charge (SoC) throughout the tour (indicated by a solid line) and the predicted 99.9\% worst case scenario (shown as a dashed line and calculated assuming normal distribution of the energy consumption along the edges). Figure \ref{fig:tour_example:b} shows the Time and SoC statistics of the tour from a 10,000 Monte-Carlo runs. The tour was calculated such that the probability of the SoC to be less than 0 along the tour is less than $0.1\%$, and the cost function is to minimize the $90\%$ worst case scenario of the tour time.

\section{Problem Statement}
The problem is described with a fully weighted connected graph $G = (V,E)$, where $V = (D, C, S)$ is the set of nodes which represent the depots $D = \{d_1,...,d_{n_d}\}$, the customers $C = \{c_1,...,c_{n_c}\}$ and charging stations $S$\footnote{$S$ may contain duplicate virtual charging station to allow more than a single visit in the same charging station} = $\{s_1,...,s_{n_s}\}$ . Every depot has a known number of vehicles $m_i$ for $i\in D$, where $M = \sum_{i \in D}{m_i}$ is the total number of vehicles $M$. Each edge on the graph that connects 2 nodes belongs to a set $E = \{(i,j), \forall i,j \in V, i \neq j \}$. Every tour has to start from and end to the same depot. The maximum number of  customers any vehicle can visit along the tour is defined by $N_{max} \in \mathbf{N}$. Every customer has to be visited only once. Every charging station can be visited as many times as needed. Traveling along an edge $(i,j) \in E$ requires an associated stochastic and non-negative time and energy consumption. In this paper we choose a normal distribution for each edge to define the mean ($T^\mu_{i,j},E^\mu_{i,j}$) and variance ($T^\sigma_{i,j},E^\sigma_{i,j}$) for both time and energy associated to each edge. Travelling along any edge is independent on other edges. Every customer has a demand to be fulfilled, $q_i$ for $i \in C$, and the sum of demands along any tour has to be less or equal to the capacity $Q$ of the vehicle. Each vehicle starts the tour with fully charged battery and it has to complete the tour with a probability greater than the design goal $P_E$ - meaning, the SoC will  be less the minimum allowed at any point along the tour with probability smaller that $P_E$. Each vehicle may visit any charging station as many times as needed and charge its battery to any SoC up to the max SoC. Any time spent in a charging station is part of the tour time. The objective function is to find the tour which minimizes the probabilistic cost function with a probability of $P_T$. For example, if $P_T = 0.9$ the goal is to choose the tour that it's $90\%$ worst time is minimized.

\section{Mixed Integer Program Formulation of Robust ECVRP} \label{MIP}
The ECVRP can be described as a Stochastic  Mixed Integer Program (MIP) and solved by transforming it into a deterministic MIP and then using a Mixed Integer solver. To formulate the MIP - First, define the decision variable matrix $X \in \mathbf{B}^{NxN}$. Every term  in the matrix, $x_{i,j}$ is a Boolean variable which is 1 if a tour goes from node i to j, and 0 otherwise. The first constraint of the MIP is that the number of exits leaving depot i is as the number of vehicles in this depot:
\begin{equation} \label{eq1}
    \sum_{j \in (C,S)} x_{i,j} = m_i, \forall i \in D
\end{equation}

The second constraint ensures that every customer has single entrance:
\begin{equation} \label{eq2}
\begin{split}
    &\sum_{j \in C, i\neq j} x_{i,j} = 1, \quad \forall i \in V \\
\end{split}
\end{equation}

The third set of constraints ensures that every charging station has no more than a single entrance:

\begin{equation} \label{eq3}
\begin{split}
    &\sum_{j \in S, i\neq j} x_{i,j} <= 1, \quad \forall i \in V \\
\end{split}
\end{equation}

To ensure that every node has the same number of exits as entrances and eliminate self loop:
\begin{equation} \label{eq4}
\begin{split}
    &x_{i,i}=0, \quad i \in V \\
    \sum_{j \in V, i\neq j}{x_{i,j}} = &\sum_{j \in V, i\neq j}{x_{j,i}} , \quad \forall i \in V
\end{split}
\end{equation}

Constraints \eqref{eq1} to \eqref{eq4} are not enough to define a valid tour. Sub-tour elimination constraints are needed. There are 2 main types of sub-tour elimination types. The first one is known as the Miller–Tucker–Zemlin (MTZ) formulation, and the second one is the Dantzig–Fulkerson–Johnson (DFJ) formulation. Both formulations are valid but each has it's own pros and cons. The MTZ formulation is defined by $\mathcal{O}(N^2)$ constraints on a $N$ auxiliary variable $u_i \in \mathbf{R}$:

\begin{subequations} \label{eq:subtour}
    \begin{alignat}{1}
        &u_i = 1 + i\cdot(N_{max}+1), \quad i \in D \label{eq:subtour:a}\\
        &u_j \geq u_i + 1 - U_{max} \cdot (1-x_{i,j}),\quad i \in V,j \in (C,S), \quad i \neq j \label{eq:subtour:b}\\
        &u_i \geq 1+j\cdot(1+N_{max}) - U_{max}\cdot(1-x_{1,j}), \quad i \in (C,S), j \in D \label{eq:subtour:c}\\
        &u_i \leq (1+j)\cdot(1+N_{max}) + U_{max}\cdot(1-x_{1,j}), \quad i \in (C,S), j \in D \label{eq:subtour:d}
    \end{alignat}
\end{subequations}

Where $U_{max}$ is a large number that makes \eqref{eq:subtour} trivial if $x_{i,j}=0$. The concept of this formulation is that $u_i$ is increased every time a node is visited. When $x_{i,j}=0$ the constraint becomes trivial, when $x_{i,j}=1$ the constraint makes sure the tour is Hamiltonian. \eqref{eq:subtour:a} initializes the values for all vehicles leaving node $i$, \eqref{eq:subtour:b} describes the dynamics of $u_i$, \eqref{eq:subtour:c} and \eqref{eq:subtour:d} make sure that the returning vehicle originated from the same depot.
The MTZ formulation is a practical formulation to use for large scale problem. This formulation is essentially a integrator that increases by 1 every time a node is visited along the tour. This technique will be used later again.
The Second formulation (DFJ), makes the sub-tour eliminations in more direct approach. For any unwanted sub-tour, there is a constraint that eliminates it. There are 2 types of tours eliminations. The first, any tour that doesn't start from the depot is eliminated. For example, the tour $(4)-(5)-(6)-(4)$ should be eliminated (if, node 4 is not a depot) and it is done by:

\begin{equation} \label{eq5}
\begin{split}
    &x_{4,5}+x_{5,6}+x_{6,4} < 3
\end{split}
\end{equation}

These set of constraints are, in general, written as:
\begin{equation} \label{eq6}
\begin{split}
    \sum_{i \in K}{\sum_{j \ne i, j \in K}{x_{ij}}} < |K|-1, \quad \forall K \subsetneq \{C,S\} \quad 2\leq |K|\leq N/2
\end{split}
\end{equation}

The second type is to exclude any tour that is longer than allowed:
\begin{equation} \label{eq7}
\begin{split}
    \sum_{i \in K}{\sum_{j \ne i, j \in K}{x_{ij}}} < |K|-1, \quad \forall K \subseteq V, \quad N_{max}+1 < |K|
\end{split}
\end{equation}

The DFJ formulation that is defined by (\ref{eq6})-(\ref{eq7}) is a stronger formulation than the MTZ for this problem, but it's also exponential in size and therefore impractical to used for large scale problem. In this paper the MTZ formulation is used.

The constraints set discussed next deal with the energy constraints of each vehicle.
We define the auxiliary variables $\{\epsilon_i\}_{i=0}^{N}$ and $\{e_i\}_{i=0}^{N}$ as the battery SoC when entering and exiting node $i$, respectively. $\epsilon_i=e_i$ for any node that is not a charging station and $\epsilon_i \leq e_i$ for $i \in S$. 
We impose the following constrains on the problem:
\begin{equation} 
\label{eq8}
\probP(\epsilon_i \geq SOC_{min}) \geq P_E \quad \forall i \in V,
\end{equation}
where $SOC_{min}$ is the minimal SOC allowed.
Since, by assumption, the energy consumption of travelling on any edge is a normally distributed variable and independent on other edges, the total energy consumption to travel up until some node along the tour is a sum of independent normally distributed variables and therefore is also a normally distributed with the following properties: 
\begin{equation} 
\label{eq9}
\epsilon_k \sim \mathcal{N}(\mu=\sum_{i \in W^S_k}{\Delta E_i} + \sum_{(i,j) \in W_k}{E^\mu_{i,j}}, \quad \sigma^2= \sum_{(i,j) \in W_k}{(E^\sigma_{i,j})^2})\
\end{equation}
Where $W_k$ is the set of edges that lead from the depot to node $k$ - a section of the tour, $W^S_k$ is the set of charging stations that were visited from the depot up until node $k$ and $\Delta E_i$ is the charged energy in node $i$. We use the MTZ approach to transform \eqref{eq8} into a set of inequalities. The MTZ approach allows for the summation of the edges variances along the tour, which produces the following set of constraints:

\begin{subequations} 
\label{eq:enegy}
    \begin{alignat}{1}
        &\epsilon_j = SOC_0, \quad \forall j \in D \label{eq:enegy:a}\\
        &e^\sigma_j = 0, \quad \forall j \in D \label{eq:enegy:b}\\
        &ef^\sigma_j = 0, \quad \forall j \in D \label{eq:enegy:c}\\
        &\epsilon_j \geq e_i + E^\mu_{i.j} - E_{max} \cdot (1-x_{i,j}),\quad i,j \in V, \quad i \neq j \label{eq:enegy:d}\\
        &(e^\sigma_j)^2 \geq (e^\sigma_i)^2 + (E^\sigma_{i,j})^2 - (x_{i,j}-1) \cdot E_{max},\quad i,j \in V \quad i \neq j \label{eq:enegy:e}\\
        &(ef^\sigma_i)^2 \geq (e^\sigma_i)^2 + (E^\sigma_{i,j})^2 - (x_{i,j}-1) \cdot E_{max},\quad i\in(C,S),j \in D  \label{eq:enegy:f}\\
        &e_i = \epsilon_i, \quad i \in \{D,C\} \label{eq:enegy:g}\\
        &e_i -\Phi(P_E) \cdot e^\sigma_i \geq  SOC_{min},  \quad i \in (C,S)\label{eq:enegy:h}\\
        &e_i - E^\mu_{i,j} - \Phi(P_E) \cdot ef^\sigma_i \geq SOC_{min} - E_{max}\cdot(1-x_{i,j}),  \quad i\in(C,S),j \in D \label{eq:enegy:i}
    \end{alignat}
\end{subequations}
Where $E_{max}$ is a large number that makes \eqref{eq:enegy:d}-\eqref{eq:enegy:i} trivial if $x_{i,j}=0$. $\{e^\sigma_i| i \in V\}$ is the standard deviation of the energy consumption from the depot to node $i$ - it's the sum of the energy variances of all edges on section of the tour from the depot to node $i$, $\{ef^\sigma_i| i \in V\}$ represent the standard deviation of the energy consumption along the complete after returning to the depot from node $i$ (it only has a meaning, if $x_{i,j}=1, j \in D$) and $\Phi(\cdot)$ is the standard normal CDF inverse function. Equations \eqref{eq:enegy:a}-\eqref{eq:enegy:c} are variable initialization for the energy and energy consumption standard deviation at the depots. Equation \eqref{eq:enegy:d} is the mean SoC change along the tour when moving along edges from the depot to node $i$. Equation \eqref{eq:enegy:e} is the summation of the energy consumption covariance along the tour when moving along edges from the depot to node $i$. Equation \eqref{eq:enegy:f} is the energy consumption covariance along the tour up to the depot. Equation \eqref{eq:enegy:g} represents the fact that the SoC when entering and exiting a non charge station node are equal. Equations \eqref{eq:enegy:h} enforces the SoC when entering node $i$ to be more than $SOC_{min}$. \eqref{eq:enegy:i} enforces the SoC to be such that when returning to depot the SoC would be more than $SOC_{min}$ when coming back to the depot.
Proving that \eqref{eq:enegy} is equivalent to \eqref{eq8} is straightforward and can be done by using the fact that a normally distributed variable can be upper bounded by knowing it's mean and variance, meaning for any randomly distributed variable $X\sim \mathcal{N}(\mu,\sigma^2)$:

\begin{subequations} \label{eq:random}
    \begin{alignat}{1}
        &\probP( X \geq \alpha) \leq P_L \iff \mu - \Phi(P_L)\cdot\sigma \geq \alpha \label{eq:random:b} \\
        &\probP( X \leq \alpha) \geq P_U \iff \mu + \Phi(P_U)\cdot\sigma \leq \alpha \label{eq:random:a}
    \end{alignat}
\end{subequations}
Where $\Phi(\cdot)$ is the normal distribution inverse CDF function. Proof for this can be found in \cite{WitteWitte}. Constraint \eqref{eq:enegy:h} ensures that the SoC at each node is more than $SOC_{min}$ by \eqref{eq:random:b} and \eqref{eq:enegy:i} ensures that the SoC when returning to the depot is more than $SOC_{min}$ by \eqref{eq:random:b}.

The connection between the entering and exiting SoC at a charging station node depends on the charging model. For a linear charging model:
\begin{equation} 
\label{eq:charging}
\begin{split}
    &e_i = \epsilon_i + \tau_i \cdot c, \quad i \in S \\
    &\tau_i \geq 0, \quad i \in V
\end{split}
\end{equation}
Where $\{\tau_i\}_{i=0}^n$ is the charging time at each node. This formulation \eqref{eq:enegy}-\eqref{eq:charging} is essentially an integrator on the SoC state of each vehicle. The charging model can be written as a more realistic nonlinear charging model where the charging rate is SoC dependent, as a piece-wise linear function presented in \cite{MONTOYA201787}.

The last set of constraints define the load capacity. Each customer has a some known demand and each vehicle has a known maximal capacity. The sum of demands of all customers visited by the same vehicle need to be less than the vehicle's maximal capacity. This can be guaranteed by summing the demand along every vehicle tour and constraining the sum to be less than allowed. As before, this can be achieved by using the impractical DFJ formulation or the MTZ formulation. Introducing the variable $\{c_i\| i \in V\}$ which represent the sum of demand from the depot to node i along the tour:

\begin{subequations} \label{eq:capacity}
    \begin{alignat}{1}
        &c_i = 0 , i \in D \label{eq:capacity:a} \\
        &c_j \geq c_i + Q_j - Q_{max} \cdot x_{i,j}, \quad i \neq j,\quad i,j \in V, i \neq j \label{eq:capacity:b} \\
        &q_i <= Q, i \in C \label{eq:capacity:c} 
    \end{alignat}
\end{subequations}
Where, $Q_{max}$ is a large number that makes \eqref{eq:capacity:b} trivial. \eqref{eq:capacity:a} is the initialization of the dispatched load in the depots, \eqref{eq:capacity:b} describes the accumulated dispatched load after visiting node i and \eqref{eq:capacity:c} constraints the dispatched load to be less than the vehicle capacity.

We use a cost function which minimizes the sum of each vehicle tour time,including the travel and charging time, next denoted as `tour total time'. Since, the travelling time along any edge is normally distributed and independent of all other edges, the total travelling time is therefore also a normally distributed variable with the following properties:
\begin{equation} 
\label{eq:TourTime}
T \sim \mathcal{N}(T_\mu=\sum_{i \in S}{\tau_i} +\sum_{i,j \in V, i \neq j}{x_{i,j} \cdot T^\mu_{i,j}}, \quad T_\sigma=\sqrt{\sum_{i,j \in V, i \neq j}{(x_{i,j}} \cdot T^\sigma_{i,j})^2})
\end{equation}

We minimize the worst case tour total time with probability $P_T$ and use again \eqref{eq:random:a} for this purpose. The cost function is therefore formulated as:
\begin{equation} 
\label{eq:CostFun}
\min_{X, \epsilon, e,\epsilon, ef,e^\sigma,ef^\sigma, \tau}{\sum_{i \in S}{\tau_i} + \sum_{i,j \in V, i \neq j}{x_{i,j} \cdot T^\mu_{i,j}} + \Phi(P_T)\cdot\sqrt{\sum_{i,j \in V, i \neq j}{(x_{i,j} \cdot T^\sigma_{i,j})^2}}}
\end{equation}
Where $\Phi(\cdot)$ is the value of the CDF function of a normal distribution. $\{\tau_i | i \in S\}$ is the charging time at the charging stations, $\{x_{i,j}|i,j \in V\}$ is the decision variable.

Since the cost function and constraints are non linear the program of the RECVRP is a Non-Linear Mixed Integer Problem (NLMIP). The NLMIP represented by \eqref{eq1}-\eqref{eq:subtour},\eqref{eq:enegy},\eqref{eq:capacity},\eqref{eq:CostFun} can be solved by a few solvers and one of them is  Gurobi~\cite{gurobi}. It is very difficult to solve this program for large instances, but it can be used to generate benchmark results for small problem instances.

\section{Clustering Heuristics}
\label{Cluster}
 The robust ECVRP \eqref{eq1}-\eqref{eq:subtour},\eqref{eq:enegy},\eqref{eq:capacity},\eqref{eq:CostFun} is an NP-hard problem.  For fast solution times, heuristic methods can be introduced. In this paper a clustering heuristic is proposed. The main idea is to divide the customers nodes to $M$ groups, and than each group will be served by a single vehicle. This reduces the big RECVRP to $M$ smaller RECTSP problems that can be solved by using various methods (further discussion in section \ref{RECTSP}). Choosing the clustering method is non trivial \cite{article3}. Attempts to solve the VRP with clustering heuristics include the work in \cite{cluster,VIDAL201587,article4}. Since the ECTSP is smaller in size compared to the original ECVRP, it's optimal or suboptimal solution can be found faster with any existing technique. The main idea in clustering the nodes is the understanding that a single vehicle should serve nodes that are close to each other - this  assumption is likely to be more applicable 
  as the size of the problem increase. When dividing the nodes into groups, we split the  large problem matrices (with size $\mathbf{R}^{NxN}$) into smaller matrices, and if the clustering is done wisely the smaller matrices might not be ill-conditioned (for example, the 2 most distant nodes would probably end up in different groups and the edge between them will not be considered). Each group should include the depot node and each group's total customers demand should be less than the maximal capacity of a single vehicle. It is also possible to limit the number of customers in each group to the max number a single truck can visit in a tour - $N_{max}$. This leads to the constrained clustering problem - divide the graph into groups that minimize some cost function, while each group meets a set of defined constraints (in this case, load and group size).
  
Defining a  distance measure and how to divide the graph into, so that the solving the smaller problems results into a solution close to the optimal solution is again an NP hard problem. In this paper we examine a few techniques and analyze the pros and cons of each method. Next we formulate the clustering problem. The nodes in $C$ should be divided into $M$ groups such that some cost function will be minimized. Each group should be associated to a vehicle and contain the depot node from which the vehicle starts and ends its tour. The matrix describing the cost of edges on the graph is given. Since the cost between nodes is not always linear with distance (edge connecting 2 nodes with an obstacle between them), it's not always possible to calculate a new cost matrix or add new auxiliary nodes (K-mean method would no be possible to implement with available data). The charging station nodes will be added after the clustering process is completed to include only the relevant charging station nodes to each group. The formulation for the clustering problem is as follows:
\begin{subequations}  \label{eq:cluster}
    \begin{alignat}{1}
        &\min_{G_i, i=1,...,M} {\sum_{i = 0}^{m}{f(Z_i)}} \label{eq:cluster:a}\\
        &s.t. \quad |G_i|<=N_{max}, \quad \forall i=1,...,M \label{eq:cluster:b}\\
        &\sum_{j \in G_i}{q_j} \leq Q, \quad \forall i=1,...,M \label{eq:cluster:c} \\
        & G_i \bigcap G_j \subseteq \{D\}, \quad i,j=1,...,M, i \neq j \label{eq:cluster:d}\\
        & |G_i \bigcap D| = 1, \quad i=1,...,M, \label{eq:cluster:e}\\
        & \sum_{i=1,...,M} {|G_i \bigcap D_j|} = m_j, \quad j \in D, \label{eq:cluster:f}\\
        & \bigcup_{i=1,...M}G_i = \{D,C\}  \label{eq:cluster:g}
    \end{alignat}
\end{subequations}
 In~(\ref{eq:cluster:g}), $G_i$ is the set of all nodes in the i-th group, $f(\cdot)$ is some function that takes a square matrix as input and outputs a scalar, $Z_i \in \mathbf{R}^{N_ixN_i}$ is the $i$-th group cost matrix which represents the cost of travelling along any edge in the group. \eqref{eq:cluster:b} limits the size of each group, \eqref{eq:cluster:c} makes sure that the sum of demand in each group is less than the capacity of each vehicle, \eqref{eq:cluster:d} ensures that the intersection between any 2 groups can only be at most a single depot node (this can only be if both group start at the same depot), \eqref{eq:cluster:e} ensures that each group will have a single depot, \eqref{eq:cluster:f} ensures that the number of vehicles associated to depot $j$ is as defined in the problem formulation - $m_j$. \eqref{eq:cluster:g} ensures that the union of all groups is the set of all depots and customers. 
 The cost matrix used in this paper is:
 \begin{equation} 
\label{eq19}
Z = T_{\mu} + \Phi(P_T) \cdot T_{\sigma} + \frac{1}{R}\cdot (E_{\mu} + \Phi(P_E) \cdot E_\sigma),
\end{equation}
where $Z \in \mathbf{R}^{{N}\times{N}}$ and $R$ is a tuning parameter linked to the average charging rate of the charging profile. This cost matrix takes into account the time of travel and energy consumption.

In case of spatial clustering, the cost function to be minimized should represent the euclidean distance between the points in the group. The cost function $f(\cdot)$, should be invariant to the order the nodes are arranged in the cost matrix $Z_i$. The cost function we testes in this paper are: 1) The Frobenius norm of the cost matrix - this cost function is easy to calculate (time complexity - $\mathcal{O}(N^2)$ ) but is dominated by the large numbers in the cost matrix that are usually not part of the solution, 2) The absolute value of the eigenvalues of the cost matrix  - calculation of all the eigenvalues of large matrices can be time consuming (time complexity - $\mathcal{O}(N^3)$), 3) The max of the absolute value eigenvalue of the cost matrix multiplied the the group size which can be efficiently calculated (time complexity - $\mathcal{O}(N^2)$ ).
 

The algorithm used in this paper to cluster the groups is a simple and straight forward greedy search algorithm - it switches nodes between groups to lower the cost function until convergence or a maximum number of iterations is reached. The clustering algorithm is described in detailed  Algorithm \ref{alg:Cluster}. After the clustering of the customers into groups is completed, each group is added a number of charging stations - only the relevant stations for each group are selected. This process simple and is  described in Algorithm \ref{alg:add_CS}.

\begin{algorithm} [h]
{\footnotesize
    \caption{Clustering Algorithm For ECVRP with multiple depots}
    \label{alg:Cluster}
    \begin{algorithmic}[1]
    \Require $D, C, Demand, CostMatrix$ \Comment{input: depot and customers groups, demand vector and cost matrix}
    \Ensure $Groups$ \Comment{return: the nodes divided into groups}

    \For{$i$ in range($M$)} \Comment{initialize the relevant depot to each group}
        \State $Groups[i] \leftarrow \emptyset$
        \State $Group[i] \leftarrow d_i$ \Comment{$d_i \in D $, is the depot node from which car i exits}
    \EndFor
    \For{$i$ in range($|C|$)} \Comment{initialize each customer to its cost minimizer, if load allows}
        \For{$j$ in range($M$)}
            \State iCost[$j$] = CalcCost([Group[$j$],$i$]) \Comment{CalcCost($\cdot$) return cost of the group}
        \EndFor
        
        \For{$j$ in argsort(iCost)} \Comment{For loop on groups with ascending cost function}
            \If{sum(Demand(Group[$j$]))+Demand[$i$] < $Q$} \Comment{If demand allows, add to group}
                \State $Group[j] \leftarrow i$
                \State break
            \EndIf
            \If{$j$ == $M$}
                \State return [] \Comment{Can't find feasible solution. add more vehicles.}
            \EndIf
        \EndFor
    \EndFor
    \State $PrevTotalCost \leftarrow \infty$
    \State $TotalCost = CalcTotalCost(Groups)$ \Comment{$CalcTotalCost(\cdot)$ calculate the total cost of current groups solution}
    \While{$TotalCost < PrevTotalCost$}
        \State $PrevTotalCost = TotalCost$
        \For{$i$ in range($|C|$)} \Comment{Insert each customer in the group that minimizes Cost}
            \State $iGroup = FindGroupOfNode(Groups,C_i)$  \Comment{find group that contains $C_i$}
            \State $Groups[iGroup].remove(C_i)$ \Comment{remove $C_i$ from it's current group}
            \State $argMinGroup = CalcCostReplaceNode(Groups, C_i)$ \Comment{calculate the group that minimizes cost when $C_i$ is added to it and feasible}
            \State $Groups[argMinGroup] \leftarrow C_i)$
        \EndFor
        \State
        \For{$i$ in range($M$)} \Comment{Try to switch Customers between groups with 1/1 ratio, only if feasible}
            \For{$j$ in range($i,M$)}
                \For{(k,l) in premuts(Group[i],Group[j],1,1)} \Comment{premuts(G1,G2,i,j) returns all unique permutations with i terms from G1 and j terms from G2}
                    \State Group[i], Group[j] = CalcGroupsSwitch(Group[i], Group[j],$C_k$,$C_l$) \Comment{CalcGroupsSwitch return feasible groups that minimize cost by switching node $k$ and $l$}
                \EndFor
            \EndFor
        \EndFor
        \State
        \For{$i$ in range($M$)} \Comment{Try to switch Customers between groups with 2/1 ratio, only if feasible}
            \For{$j$ in range($M$)}
                \For{(k1,k2,l) in premuts(Group[i],Group[j],2,1)} 
                    \State Group[i], Group[j] = CalcGroupsSwitch21(Group[i], Group[j],$C_{k1}$,$C_{k2}$,$C_l$) \Comment{CalcGroupsSwitch return feasible groups that minimize cost by switching node $k_1$ and $k_2$ with $l$}
                \EndFor
            \EndFor
        \EndFor
        \State TotalCost = CalcTotalCost(Groups)
    \EndWhile
    \State return Groups
    \end{algorithmic}
}
\end{algorithm}

\begin{algorithm} [h]
{\footnotesize
    \caption{Clustering Algorithm For ECVRP with multiple depots}
    \label{alg:add_CS}
    \begin{algorithmic}[1]
    \Require $Groups, S, CostMatrix$ \Comment{input: Clustered groups, Charging Stations nodes and cost matrix}
    \Ensure $Groups$ \Comment{return: groups with charging stations}
        \For{$i$ in range($M$)}
            \For{iNode in Group[i]}
                \State $iClosestCS = argmin(CostMatrix[iNode,CSs])$ \Comment{CSs is group of all charging stations}
                \If{$CSs[iClosestCS] not in Group[i]$}
                    \State $Group[i] \leftarrow CSs[iClosestCS]$
                \EndIf
            \EndFor
        \EndFor
    \end{algorithmic}
}
\end{algorithm}

\section{Solving The Robust ECTSP} \label{RECTSP}


After the large Robust ECVRP has been divided into an number of smaller Robust ECTSP problems, a numerical solver can be used to solve for the optimal tour for each vehicle. This can be done by solving a MIP, any of the state-of-the-art ECTSP algorithms, by a simple search algorithm like Dynamic Programming (DP) or a  Divide and Conquer (D\&C) method. Since the ECTSP problem may still be large, we can use the same clustering algorithm that was described in Section \ref{Cluster} to reduce the dimension of the problem,i.e., divide the nodes in the group to subgroups such that moving from subgroup to subgroup is possible only if all the node in the subgroup has already been visited. 
This is the approach we take in this paper.
Furthermore, due to the additional energy constraints, many trajectories can be eliminated early in the search. Note that the load capacity problem is irrelevant at this point, since the clustering of the groups made sure that the capacity constraint is met. The number of subgroups can be determined by tuning or to ensure fast computation times - The lower the number of subgroups the faster the search algorithm will find a solution. The higher the number of subgroups, it's more likely the solution will get close to the optimal Robust ECTSP tour. The main idea of the solver is to search recursively all possible routes, while stopping the search whenever the current best potential route is worst than the best route found so far, or it's infeasible due to energy constraints. The search is done using the greedy method to sort the order of search to speed up finding good feasible solution and eliminate many potential tours as early as possible. The cost of each feasible route is calculated according to the charging profile with the optimal charging times to find the optimal charging profile. This recursive algorithm can be parallelized in a multi core processor with a shared memory. The algorithm can be stopped with a timeout condition if some feasible solution is found. The algorithm is described in details in Algorithm \ref{alg:search}.

\begin{algorithm} [h]
{\tiny
    \caption{ECTSP solving algorithm}
    \label{alg:search}
    \begin{algorithmic}[1]
    \Require $Depot, ChargingStations, SubGroups, ScenarioData$ \Comment{input: depot and charging stations, and Sub-Groups of customers nodes, mean and covariance time/energy matrices}
    \Ensure $GroupRoute, Cost, ChargingProfile$ \Comment{return: Vehicle Route, Cost of route and charging profile}
    \State Function $Route, Cost, ChargingProfile =$  \\ \qquad\qquad$ECTSP\_solver(CurNode, CurRoute, CurTime, CurSOC, CurTimeVar, CurEnergyVar, BestRoute, BestCost, ChargingPotential)$
        \State $CurRoute \leftarrow CurNode$
        \If{$AllNodesVisited(CurRoute) == True$} \Comment{Check if traveling to depot is feasible}
            \State $CurRoute \leftarrow depot$
            \State $CurTime, CurSOC, TimeVar, EnergyVar = $ \\ \qquad\qquad $UpdateRouteParams(CurTime, CurSOC, TimeVar, EnergyVar, depot)$
            \If{$CurSOC + ChargingPotential - Phi_E*sqrt(EnergyVar) < 0$}
                \State return $CurRoute, inf, []$
            \Else
            \State $ChargingProfile = CalcOptimalChargingProfile(CurRoute, ScenarioData)$ \Comment{Calculate optimal charging profile according to visited charging stations and charging profile}
            \State $CurCost = CalcCost(CurRoute, ChargingProfile)$
                \State return $CurRoute, CurCost, ChargingProfile$
            \EndIf
        \EndIf
        \State
        \If {$CurNode in ChargeStationsNodes$} \Comment{If current node is a charging station}
            \State $ChargingPotential = SOC_{max} - CurSOC$
        \EndIf

        \State $NextPossibleNodes = NodesInSubGroup(CurNode)$ \Comment{This function return the nodes in the subgroup of $CurNode$}
        \State $NextPossibleNodes - set{NextPossibleNodes} - set{CurRoute}$
        \If{$NextPossibleNodes == []$} \Comment{If all nodes in subgroup were visited - move to next subgroup}
            \State $NextPossibleNodes = set{allNodes} - set{CurRoute}$
            \State $SortedNodes = sort(TimeMatrix[CurNode,NextPossibleNodes]$ \Comment{sort nodes by distance from $CurNode$}
        \EndIf
        \For{iNode in SortedNodes}
            \State $NextTime, NextSOC, NextTimeVar, NextEnergyVar = UpdateRouteParams(CurTime, CurSOC, TimeVar, EnergyVar, CurNode, iNode)$
            \If {$NextTime + Phi_T*sqrt(TimeVar) \geq BestCost$}
                \State continue
            \EndIf
            \State $MinPossibleTimeToCompleteTour = MinTimeEstimate(CurTime, NodeLeftToVisit)$
            \State $MinPossibleTimeVarToCompleteTour = MinTimeVarEstimate(TimeVar, NodeLeftToVisit)$
            \State $MinPossibleEnergyToCompleteTour = MinEnergyEstimate(CurEnergy, NodeLeftToVisit)$
            \State $MinPossibleEnergyVarToCompleteTour = MinEnergyVarEstimate(EnergyVar, NodeLeftToVisit)$
            \If{$MinPossibleTimeToCompleteTour + Phi_T*sqrt(MinPossibleTimeVarToCompleteTour) \geq BestCost$}
                \State continue
            \EndIf
            \If{$MinPossibleEnergyToCompleteTour + Phi\_E*sqrt(MinPossibleEnergyVarToCompleteTour) + MinPossibleChargingPotential < 0$}
                \State continue
            \EndIf
            \State $NextTime, NextSOC, NextTimeVar, NextEnergyVar = UpdateRouteParams(CurTime, CurSOC, TimeVar, EnergyVar, iNode)$
            \State $Route, Cost, ChargingProfile = $\\ \qquad $RECTSP\_solver(iNode, CurRoute, NextTime, NextSOC, NextTimeVar, NextEnergyVar, BestRoute, BestCost, ChargingPotential)$
            \If{$Cost < BestCost and Route[-1]==depot$}
                \State $BestCost = Cost$
                \State $BestRoute = Route$
                \State $BestChargingProfile = ChargingProfile$
            \EndIf
        \EndFor
        \State return $BestRoute, BestCost, BestChargingProfile$
    \end{algorithmic}
}
\end{algorithm}

\section{Results}
This section presents the tuning of the clustering algorithm cost function and the results of solving the Robust ECVRP using the presented methodology on two set of examples. The first set (set 1) is a randomly generated set of scenarios - in each scenario the total number of nodes and vehicles are given. The position of all nodes are randomized from uniform distribution on a $100 \times 100$ map. Each Customer node demand is randomized from uniform distribution $q_i \sim \mathcal{U}[5,20]$. The maximal vehicle load capacity is $Q=100$. The energy consumption is linear function of the travel distance $e_{i,j} = c \cdot r_{i,j}$, where $c = 1.85$ where the maximal battery capacity is $SOC_{max}=100.0$. The charging profile is linear with a charging rate of $3$. The standard deviation of the time and energy is randomized at $5\%$ to $30\%$ of the nominal value for each edge. 10\% of the nodes are randomized to be a charging stations. 

The second set (set 2) of examples is  the well-known  data-set proposed in \cite{EVRP_dataset}. This is a data-set for a small (up to 100 nodes) and large (up to 1000 nodes) ECVRP instances with many attempts of solving them. To turn these instances to stochastic instances, the variance of the time and energy consumption is assumed to be 1\% from the nominal value. The probability parameters used are $P_E = 0.999$ for the energy constraints and $P_T = 0.9$ for the cost function. The results are compared to the Best Known Solution (BKS) for each instance. We compare 2 main parameters: the quality of the solution (how close can it get to the BKS) and the time of calculation. The results presented for the deterministic and stochastic cases. All resulted calculated for this paper have been done on a Laptop equipped with Intel® Core™ i7-1260P 3.4 GHz and 16Gib of RAM. All code used in this paper was implemented in Python 3.11.5 that runs on Ubuntu 22.04 OS. Note that if the same algorithm were to be implemented in C/C++ (like most other algorithms compared) the computation times presented in the results section (table \ref{tab:results_random} and \ref{tab:results_large}), will be reduced.

\subsection{Tuning of the cost function of the clustering algorithm}

In this section, the cost function type for the clustering algorithm is explored. The clustering algorithm with different cost function is used to cluster the instances using algorithm \ref{alg:Cluster} and then the problem is solved using algorithm \ref{alg:search}.
In order to find the best tuning, we used the randomized scenarios and the small data-set of \cite{EVRP_dataset} (data-set E). For these small data-sets a solution to RECVRP can be found using a MIP solver (the clustering based solution is used as initial guess for the MIP solver). 
The results are used to choose the best type of cost function for the clustering algorithm. Three are the costs considered next. Cost \eqref{eq:costTypes:a} takes the sum of the maximal eigenvalue of the cost matrix multiplied by the number of nodes in the group (this is done using the Power Method for numerical efficiency). Cost \eqref{eq:costTypes:b} is the sum of the Frobenius norm of all the cost matrices. Cost \eqref{eq:costTypes:c} is the sum of the absolute value of all the eigenvalues of the cost matrices:
\begin{subequations}  \label{eq:costTypes}
    \begin{alignat}{1}
        &f_1 = \sum_{i=1}^M{\lambda_{max}(Z_i) \cdot |G_i|}  \label{eq:costTypes:a}\\
        &f_2 = \sum_{i=1}^M{|Z_i|_F} \label{eq:costTypes:b}\\
        &f_3 = \sum_{i=1}^M{\sum_{j=1}^{|G_i|}{|\lambda_j(Z_i)|}} \label{eq:costTypes:c} 
    \end{alignat}
\end{subequations}
In all costs $Z_i$ is the cost matrix of group $i$  and $G_i$ is the set of all nodes in group $i$.
Table \ref{tab:results_cost} shows the results of the tuning scenarios  solved with different cost function. It shows the cost of the tours produced by solving the MIP with Gurobi solver (Gurobi was run with a timeout of 10 min), and a solution using one of the clustering cost function \eqref{eq:costTypes}. The name of the scenario suggests it's type and structure - The first letter is D if the scenario is deterministic ($T_\sigma = E_\sigma = 0$) or S for stochastic. The second letter is E (or X) if the scenario is from \cite{EVRP_dataset} or R if the scenario is randomized. The letter "n" followed by a number represent the number of customers in the scenario. The letter "k" followed by a number represent the number of vehicles in the scenario. All the scenarios in E are single depot and R are multiple depots. It can be seen that as the number of nodes increase, cost function \eqref{eq:costTypes:c} gets better and better results relative to \eqref{eq:costTypes:a} and \eqref{eq:costTypes:b}. All cost function types have similar calculation times. Based on results in table \ref{tab:results_cost}, the chosen cost function is \eqref{eq:costTypes:c} - it gives usually the best results and the computation times are similar to the other methods. From now on the results in this paper are shown using this cost function.

\begin{table} [h]
\begin{center}
\caption{Results for Algorithm Parameter Tuning}
\begin{tabular}{||c c c c c c||} 
\hline
Scenario & $MIP \  Solver/BKS$ & $MIP \ Gap [\%]$ & $f_1$ & $f_2$ & $f_3$\\ [0.5ex] 
\hline
\hline
SE-n22-k4 Cost & 400.0 & 4.0 & 421.0 & 414.1 & 411.3  \\ 
Clustering Time [s] & - & - & 0.1 & 0.1 & 0.1 \\ 
\hline
SE-n23-k3 Cost & 642.0 & 12.3 & 712.5 & 644.5 & 642.0   \\ 
Clustering Time [s] & - & - & 0.3 & 0.2 & 0.3  \\ 
\hline
SE-n33-k4 Cost & 904.8 & 11.1 & 960.9 & 915.1 & 904.8  \\ 
Clustering Time [s] & - & - & 0.2 & 0.3 & 0.7  \\ 
\hline
DE-n22-k4 Cost & 384.7 & 0.0 & 393.6 & 390.3 & 390.3  \\ 
Clustering Time [s] & -& - & 0.1 & 0.1 & 0.1  \\ 
\hline
DE-n23-k3 Cost & 571.9 & 0 & 665.5 & 665.5 & 665.5  \\ 
Clustering Time [s] & - & - & 0.3 & 0.2 & 0.3  \\ 
\hline
DE-n33-k4 Cost &  &  & 910.4 & 885.3 & 883.5  \\ 
Clustering Time [s] & - & - & 0.3 & 0.3 & 0.3  \\ 
\hline
SE-n51-k5 Cost & 558.5 & 11.4 & 586.4 & 558.5 & 590.5  \\ 
Clustering Time [s] & - & - & 0.7 & 0.9 & 1.2 \\ 
\hline
SE-n101-k8 Cost & 903.9 & 17.9 & 946.4 & 922.7 & 903.9  \\ 
Clustering Time [s] & - & - & 7.5 & 11.2 & 14.5 \\ 
\hline
SR-n30-k4 Cost & 173.6 & 18.0 & 198.0 & 167.9 & 177.4  \\ 
Clustering Time [s] & - & - & 0.4 & 0.4 & 0.4  \\ 
\hline
SR-n50-k4 Cost & 223.7 & 19.7 & 213.3 & 228.2 & 227.0   \\ 
Clustering Time [s] & - & - & 1.3 & 1.4 & 1.3 \\ 
\hline
SR-n100-k7 Cost & - & - & 353.5 & 364.8 & 360.4   \\ 
Clustering Time [s] & - & - & 2.1 & 5.4 & 4.8  \\ 
\hline
SR-n164-k18 Cost & 223.7 & 19.7 & 768.2 & 701.6 & 699.4   \\ 
Clustering Time [s] & - & - & 36.5 & 31.9 & 28.4 \\ 
\hline
\end{tabular}
\label{tab:results_cost}
\end{center}
\end{table}

\subsection{Results For Set 1}

The results for randomized small instances are shown and compared to the solution calculated using the MIP formulation described in section\ref{MIP} solved using Gurobi solver \cite{gurobi}. Each instance was solved 10 times with different seeds and the average of all 10 runs are presented. Table \ref{tab:results_random} shows the results. The reported gap of the MIP is the reported gap by the solver at the end of the run. The reported gap using the Clustering Algorithm is relative to the MIP solution (the solution with the clustering algorithm was used as initial guess for the MIP solver). It can be seen that the solution achieved by the clustering algorithm is very close to the MIP solution even for small instances. The instance name suggests it's  - all start with "SR" (Stochastic Random), n is the number of total nodes, k - represent the number of depots and the number of vehicles in each depot (k21 - there 2 depots, 2 vehicles in the first and 1 in the second).

\begin{table} [h!]
\begin{center}
\caption{Results for small Randomly Generated instances}
\begin{tabular}{||c c c c c||} 
\hline
Scenario & MIP Cost & MIP Gap [\%] & Clustering Algo. Cost [s] & Clustering Algo. Gap [\%] \\ [0.3ex] 
\hline
\hline
SR-n11-k11 & 114.2 & 0.0 & 117.2 & 2 \\ 
\hline
SR-n12-k11 & 109.6  & 0.0 & 112.5 & 2 \\ 
\hline
SR-n15-k22 & 136.6  & 0.0 & 139.4 & 2.1 \\ 
\hline
SR-n20-k22 & 161.9  & 7.4 & 171.0 & 5.4 \\ 
\hline
SR-n25-k221 & 160.1  & 3.4 & 167.5 & 4.6 \\ 
\hline
SR-n30-k32 & 161.9  & 7.4 & 171.0 & 5.4 \\ 
\hline
\end{tabular}
\label{tab:results_random}
\end{center}
\end{table}

\subsection{Results For Set 2}

 Table \ref{tab:results_large} shows the results for the instances \cite{EVRP_dataset} which are solved twice - once as a deterministic problem (starts with "D") and then as a stochastic (starts with "S"). The deterministic case is compared to the BKS in terms of quality and time of solution. The BKS and best calculation time is taken from the best solution reported in \cite{mavrovouniotis2020benchmark,woller2020grasp,9409782}. The results for the stochastic case are presented as a benchmark for future research. Table \ref{tab:results_large} shows the results of the clustering based algorithm. It can be seen that the algorithm is faster than the fastest algorithm that has calculation time statistics. For the largest instances, calculation times are 3-4 times faster than the faster algorithm reported. In terms of the quality of solution, as the instances get larger, the algorithm gets very close to the BKS. This can be explained by the intuition that for large instances the tours are more compact so that a clustering heuristics works very well. The cost presented is:
 \begin{equation} 
\label{eq20}
Cost = T_{\mu} + \Phi(P_T) \cdot T_{\sigma}
\end{equation}
Where, $P_T = 0.9$. The energy constraint are calculated with robustness probability of $P_E = 0.999$.

\begin{table} [h!]
\begin{center}
\caption{Results for Clustering Algorithm On Large Scale instances}
\begin{tabular}{||c c c c c c||} 
 \hline
 Scenario & BKS Cost & Solution Time [s] & Tour Cost & Solution Time [s] & Sol Gap [\%] \\ [0.5ex] 
 \hline
 \hline
 SX-n143-k7 & - & - & 19849 & 104 & -  \\ 
 DX-n143-k7 & 16028 & 104 & 17993 & 101 & 13.0 \\ 
 \hline
 SX-n214-k11 & - & - & 13856 & 124 & -  \\ 
 DX-n214-k11 & 11133 & 250 & 13152 & 114 & 18.0 \\ 
 \hline
 SX-n351-k40 & - & - & 13856 & 144 & -  \\ 
 DX-n351-k40 & 26478 & 500 & 28574 & 139 & 7.0 \\ 
 \hline
 SX-n459-k26 & - & - & 30547 & 635 & -  \\ 
 DX-n459-k26 & 24763 & 1900 & 28896 & 601 & 16.0 \\ 
 \hline
 SX-n573-k30 & - & - & 62067 & 1776 & -  \\ 
 DX-n573-k30 & 51929 & 2500 & 57906 & 1503 & 11.5 \\ 
 \hline
 SX-n685-k75 & - & - & 81683 & 674 & -  \\ 
 DX-n685-k75 & 70834 & 5511 & 75175 & 654 & 6.1 \\ 
 \hline
 SX-n749-k98 & - & - & 91216 & 453 & -  \\ 
 DX-n749-k98 & 80299 & 7258 & 83366 & 438 & 3.8 \\ 
 \hline
 SX-n819-k171 & - & - & 179714 & 307 & -  \\ 
 DX-n819-k171 & 164289 & 6612 & 167075 & 236 & 1.7 \\ 
 \hline
 SX-n916-k207 & - & - & 372304 & 384 & -  \\ 
 DX-n916-k207 & 341649 & 6774 & 344346 & 294 & 0.7 \\ 
 \hline
 SX-n1001-k43 & - & - & 88531 & 4965 & -  \\ 
 DX-n1001-k43 & 77476 & 8360 & 79885 & 3615 & 3.1 \\ 
 \hline
\end{tabular}
\label{tab:results_large}
\end{center}
\end{table}

\section{Conclusions}
Solving the Robust ECVRP can be done quickly with clustering heuristics that produce sub optimal solution with short computation times. The clustering heuristics is a natural and intuitive way of solving the large scale vehicle routing problem. The added complexity of the energy constraints and robustness makes the corresponding MIP extremely hard to solve, but on the other hand makes it possible for the introduction of a simple iterative search algorithm that utilized the many tours eliminated by the energy constraints. Due to the fast solving times, this algorithm can be used for a closed loop feedback loop of the whole process. 

\section{Data Availability Statement}
All the code used int this paper to calculate the results presented can be found in the repository - \UrlFont{https://github.com/pukmark/EVRP\_Optimization.git}

\bibliographystyle{IEEEtran}
\bibliography{bibliography.bib}

\end{document}